\documentclass[useAMS,usenatbib]{mn2e}

\usepackage[totalwidth=520pt, totalheight=680pt, left=15mm, letterpaper]{geometry}

\usepackage{times}
\usepackage{graphicx}

\def\beq{\begin{equation}}
\def\beqa{\begin{eqnarray}}
\def\beqaa{\begin{eqnarray*}}
\def\eeq{\end{equation}}
\def\eeqa{\end{eqnarray}}
\def\eeqaa{\end{eqnarray*}}

\newcommand{\de}{\mathrm{d}} 

\newcommand{\kmsMpc}{\;\mathrm{km}\,\mathrm{s}^{-1} \mathrm{Mpc}^{-1}}

\newcommand{\Omm}{\Omega_\mathrm{m}}
\newcommand{\Om}{\mathcal{O}_\mathrm{m}}
\newcommand{\Omz}{\mathcal{O}_\mathrm{m}^{(1)}(z)}
\newcommand{\LCDM}{$\Lambda$CDM}
\newcommand{\Ht}{$H\,t=1$ model}
\newcommand{\Rh}{R_\mathrm{h}}
\newcommand{\rh}{r_\mathrm{h}}

\begin{document}

\title[We do not live in the $\Rh=c\,t$ universe]{We do not live in the $\bmath{\Rh=c\,t}$ universe}
\author[Maciej Bilicki and Marina Seikel]{Maciej Bilicki \thanks{email:
    maciek(at)ast.uct.ac.za} and Marina Seikel \thanks{email: marina.seikel(at)uct.ac.za}\\
Astrophysics, Cosmology and Gravity Centre (ACGC), University of Cape Town, Rondebosch 7701, Republic of South Africa}
\maketitle

\begin{abstract}
We analyse the possibility that our Universe could be described by the
model recently proposed by Melia \& Shevchuk (2012), where the Hubble
scale $\Rh = c / H$ is at all times equal to the distance $c \,t$ that
light has travelled since the Big Bang. In such a model, the scale
factor is proportional to cosmic time and there is neither acceleration nor
deceleration of the expansion. We first point out problems with the
very foundations of the model and its consequences for the evolution
of the Universe. Next, we compare predictions of the model with
observational data. As probes of the expansion we use distance data of
supernovae type Ia, as well as Hubble rate data obtained from cosmic
chronometers and radial baryon acoustic oscillations. We analyse the
redshift evolution of the Hubble parameter and its redshift
derivatives, together with the so-called $\Om$ diagnostic and 
the deceleration parameter. To reliably estimate smooth functions and
their derivatives from discrete data, we use the recently developed
Gaussian Processes in Python package (GaPP). Our general conclusion is
that the discussed model is strongly disfavoured by observations,
especially at low redshifts ($z\la0.5$). In particular, it predicts
specific constant values for the deceleration parameter and for
redshift derivatives of the Hubble parameter, which is ruled out by
the data.
\end{abstract}


\section{Introduction}
The fact that the Universe is currently undergoing a phase of
accelerated expansion has been known for more than a decade. The first
evidence came from observations of distant supernovae type Ia (SNeIa)
by \citet{Riess98} and \citet{Perl99}. This result has been confirmed
by various observations, e.g.  of the cosmic microwave background and its
angular power spectrum \citep{Komatsu11}, baryon acoustic oscillations
(BAO, e.g.\ \citealt{Percival10}) and SNeIa \citep{Union2.1}; see
\citet{Weinberg} for a general review.

However, the picture that emerges based on these and multiple other
observations is far from being theoretically understood. In a
homogeneous and isotropic Universe and assuming general relativity to
be correct on cosmological scales, one needs some form of \textit{dark
  energy} as the driving force of the observed acceleration.  The
simplest model that is consistent with observations is the Lambda-Cold Dark Matter (\LCDM) --
a model where the dark energy has a constant energy density.

This theory has no physical understanding at the moment and in recent
years considerable effort has been devoted to develop and test
alternative explanations of the accelerated cosmic expansion. These
efforts include dark energy models with an equation of state that is
more complex than that of \LCDM, inhomogeneous models and theories
assuming deviations from general relativity, known under the
collective name of modified gravity [see \citet{Copeland} for a review].

\section{The $\bmath{R}$\lowercase{$\bmath{_\mathrm{h}=c\,t}$} model and its problems}
In a recent paper, \citet[][hereafter MS12]{MeSh12} have offered a
qualitatively different cosmological model, which was meant to solve some of the problems of the \LCDM.
Namely, they used the general relativistic framework of
Friedman-Lema\^{i}tre models with the Robertson-Walker metric and,
having applied the \textit{Weyl postulate}, proposed a model in which
there is neither acceleration nor deceleration either now or in the past. It is
claimed that this is the \textit{only} viable cosmological model as
all other models allegedly violate the Weyl postulate. In this
\textit{eternally coasting} model, the scale factor grows linearly
with time. Additionally, the MS12 model predicts the Hubble parameter
to be always exactly equal to the inverse of the age of the Universe,
i.e. $H(z)=t^{-1}(z)$, where $t$ is the cosmic time and $z$ is the
cosmological redshift. MS12 argue that the best-fit (concordance)
cosmological model only mimics the \textit{real} underlying one, which
they call the $R_h=c\,t$ Universe (see also \citealt{M12pop} for a
more popular description). For simplicity of notation, it will be
referred to as the `\Ht' hereunder.

As one of the motivations to introduce their model, MS12 point out
that the currently observed coincidence that $\Rh= c\, t_0$ (where
$t_0$ denotes the current age of the Universe) is possible within
$\Lambda$CDM models only once, i.e.\ exactly now. On the other hand,
as they argue, such an equality [i.e.\ $\Rh=c\slash H(t)$] is obtained
\textit{for all times} in a straightforward manner if one applies the
Weyl postulate when constructing the space-time metric.

MS12 start by defining the
gravitational radius (better known as Schwarzschild radius) as
\beq\label{eq: R_h}
\Rh\equiv\frac{2\, G M(\Rh)}{c^2}
\eeq
[equation (8) of MS12], where $M(\Rh)$ is the mass enclosed within $\Rh$
\beq
M(\Rh) = V_\mathrm{prop} \frac{\rho(t)}{c^2}\;,
\eeq
and 
\beq\label{volume}
V_\mathrm{prop} = \frac{4\pi}{3}\Rh^3
\eeq
is the proper volume, defined by MS12 such that the comoving
density of particles within it remains fixed as the Universe
expands. Here, $\rho(t)$ denotes the energy density. 

MS12 later argue that the gravitational radius is equivalent to the
Hubble radius.  We doubt that the definition of the Schwarzschild
radius makes sense on such large scales. One of the crucial
assumptions in deriving the Schwarzschild radius is a static
space-time. This condition is certainly not true when expanding
regions are considered. Nevertheless we will continue to follow the
argumentation given in MS12 and point out another problem in their
theory. This is additional to what has been described in two earlier papers dealing with some flaws of the MS12 model \citep{vOKL10,LvO12}.

Combining the Equations (\ref{eq: R_h})--(\ref{volume}) with the Friedmann equation for a flat
Universe
\beq
H^2(t) =\frac{8\pi G}{c^2}\rho(t)
\eeq
one finds
\beq
\Rh(t) =\frac{c}{H(t)}\;,
\eeq
which is the usual definition of the Hubble radius. Using $\Rh(t) =
a(t)\rh$ and $H(t) = \dot{a}(t)/a(t)$, we find for the comoving Hubble
radius $\rh$:
\beq\label{comovingRH}
\rh = \frac{c}{\dot{a}(t)}
\eeq
The next step of MS12 argumentation contains another flaw. They claim that
$\rh$ needs to be constant because `\textit{otherwise
  $V_\mathrm{prop}$ would not represent the volume within which the
  particle density is constant in the comoving frame}'. This argument
is, however, not convincing. In a homogeneous Universe, the size of a
volume -- and thus the value of $\rh$ -- does not affect the particle
density within that volume. Therefore, there is no reason to assume
that $\rh$ is constant. Consequently, it can neither be implied that
$\dot{a}$ is constant, nor that the scale factor is linear in time,
nor that the equation of state (EOS) $w = -1/3$ (see hereunder). 

\textit{If} the argumentation of MS12 were correct, that would have
some odd consequences for the evolution of the Universe, which we will
examine in the following. Being a special case of a \textit{power-law
  cosmology} (see \citealt{Kumar12} for a recent discussion, and
references therein), in which the scale factor $a(t)\propto t^\alpha$,
this model has a simple mathematical form. However, it is arguably not
more plausible from the \textit{astrophysical} point of view than the
standard \LCDM. Firstly, as can be seen in equation (\ref{comovingRH}), a
constant $\rh$ would imply that $\dot{a}$ is constant and thus the
deceleration parameter equal to zero, $q(z)\equiv 0$.  Secondly, from Equations (\ref{eq: R_h})--(\ref{volume}), it follows that $\Rh\propto
M(\Rh)\propto\rho\Rh^3$, so we get that $\rho \Rh^2 = \mathrm{const}$
at all times. In combination with $\Rh(t) = a(t)\rh$ and assuming $\rh$
to be constant, this leads directly to the following scaling for
energy density, which must hold in the \Ht: 
\beq 
[H\,t=1]\qquad \rho(t) \propto a(t)^{-2}\;.  
\eeq
It is easy to check that it is equivalent to require the EOS to be
$w=p/\rho\equiv-1/3$, as discussed in MS12.
 
Recall that according to MS12 all other homogeneous and isotropic
cosmological models are ruled out due to (incorrect, as we believe) theoretical
reasoning. These models would not even have consistent general
relativistic solution. Thus, a homogeneous dust Universe would not
have a general relativistic solution because it has $w=0$. Neither
would inflation be possible. The contents of the Universe would need
to be fine-tuned in such a way that the overall effective EOS equals
$-1/3$ at \textit{all} times. Such a high degree of fine-tuning is
very hard to achieve. Another problem is that the growth of fluctuations in the \Ht{} is
driven by a (negative) pressure term, which looks the same no matter
the perturbation length. It is thus somewhat unclear how the large-scale structure
would emerge within this model.

Finally, however, neither simplicity nor plausibility are the decisive factors
for a cosmological model to be accepted as a proper description of the
physical Universe. The condition under which we can treat the model as
valid, or at least a good approximation to reality, is that it has
successfully passed confrontations with multiple observational
data. Not all tests, applicable elsewhere, can
be currently performed within the \Ht, owing to the lack of
theoretical framework in some domains. For example, the perturbation
theory of gravitational instability, so successfully applied in
\LCDM{} (e.g.\ \citealt{Davis11,ND11,BCJM11} for recent results),
could not be used in the discussed model. As shown in MS12, the growth
of density fluctuations is qualitatively different there from the
standard cosmology and its quantitative details still await
investigation. However, as we will show below, such analyses might
already be no more than of academic interest, as the model can be
effectively falsified with other, simpler methods. Examining the
consequences of the \Ht{} for the expansion history of the Universe
and its various indicators leads us to conclude that the MS12
model is strongly disfavoured by observations.

\section{Supernova type Ia data}
One of the most important tests of the cosmic expansion history
consists in comparing redshifts and distances of far-away
objects with predictions of particular models. For example, in case of
SNeIa, the quantity used is the \textit{luminosity distance}, derived
by lightcurve fitting from observed peak luminosities and analysed
afterwards as a function of cosmological redshift.

Assuming spatial flatness, as was done in MS12,\footnote{The analysis can be generalized to non-flat models at the expense of more complicated expressions for particular quantities.} the luminosity distance in \LCDM{} models with a cosmological constant (dark energy equation of state $w=-1$) is given by
\beq
d_\mathrm{L}=\frac{c}{H_0}(1+z) \int\limits_0^z{\frac{\de \zeta}{\left[\Omm(1+\zeta)^3+\Omega_\Lambda\right]^{1/2}}}\;.
\eeq
In the \Ht, it takes the following closed-form expression:
\beq
[H\,t=1]\qquad d_\mathrm{L}=\Rh(t_0) (1+z) \ln(1+z)\;,
\eeq
where $\Rh(t_0)=c/H_0$. In principle, comparing this curve with
available data could be  used to directly verify the validity of the
model. Indeed, MS12 discuss it shortly and even admit that this form
of the luminosity distance does not fit the current sample of Type Ia
supernovae (see e.g.\ \citealt{Riess04}). They however argue that this
discrepancy is present only for redshifts of $\sim0.5$, where -- as
they claim -- different SNeIa catalogues are inconsistent with each
other. In particular, they refer to the fact that analyses made by
different authors do not yield a consistent value of the redshift
$z_\mathrm{acc}$ at which, within \LCDM, there was a transition from
past deceleration to currently observed acceleration of the universal
expansion. 

Putting these issues aside, we only mention a known caveat related to
using luminosity distances of SNeIa, or their more observer-friendly
variant, distance moduli, as direct probes of cosmic expansion. When
using the SALT light-curve fitter \citep{SALT,SALT2} some nuisance
parameters that are needed to determine the distance moduli from the
raw data, are fitted simultaneously with the cosmological
parameters. Thus, the distance muduli in the data sets are somewhat
model-dependent. While the MLCS2k2 light-curve fitter \citep{MLCS} avoids
these issues, other systematics are present (e.g.\ the reddening
parameter $R_V$ has to be fixed by hand). Thus, although both types of light-curve fitters have some systematics, these are of different nature. So when accelerated
expansion is found independently by using different fitters, this
should be a fairly robust result.

When discussing the luminosity distance, MS12 question the validity of
introducing into analyses such parameters as the deceleration
parameter (called the acceleration parameter therein). However, it is
by examining this parameter that one can put additional constraints on
the validity of the \Ht. It has been widely used for many decades as a direct probe of the cosmic expansion and it has straightforward interpretation. For example, within \LCDM{} models (also those with $\Lambda=0$), its current value is exactly equal to $q_0=\Omm/2-\Omega_\Lambda$ (e.g.\ \citealt{Peacock}).

In general, the deceleration parameter is defined as
\beq
q(z)\equiv -\frac{a\, \ddot{a}}{\dot{a}^2}=\frac{H'(z)}{H(z)}(1+z)-1\;,
\eeq
where prime denotes differentiation with respect to redshift. The Hubble parameter is defined in the standard way,
$H\equiv \dot{a}/a$,
where dot stands for time derivative. From definition then, $q(z)$ is \textit{negative} if the universal expansion is
accelerating.  In the \Ht, 
\beq\label{eq:q in Ht1}
[H\,t=1]\qquad q(z)\equiv 0
\eeq
at all times.  However,
various independent analyses of SNeIa have shown that the current
value of the deceleration parameter is significantly less then zero
\textit{and} that it is \textit{not} constant in time. This conclusion is
independent of the choice of the light-curve fitter.

In \citet{SeSch09}, a general test of accelerated expansion has been
performed using the MLCS2k2 and SALT fitters and different data
sets. In this analysis, strong evidence for a current phase of accelerated
expansion has been found. Another example is
the kinematic analysis performed by \citet{GCL09}, which does not assume a
particular underlying cosmological model. Among several
parametrisations used therein, of particular interest is the one with
constant $q(z)$. A fit to the Union Compilation
\citep{Kow08} gave $q_0=-0.34\pm0.05$ (1$\sigma$ errorbars). 
Even with the assumption of $q$ being constant, the \Ht{} is at odds
with this result. Relaxing the assumption, strongly points towards a
$q(z)$, which is changing with time and significantly less than zero
in the late Universe -- a result that is inconsistent
with the \Ht. This has been shown in numerous analyses.
Especially those including newer
observations of SNeIa at redshifts bigger than 1, as well as other
probes of cosmic expansion (chronometers, baryon acoustic oscillations) unambiguously point to
a transition from past deceleration to current acceleration (see
e.g.\ \citealt{SCA11,Lu11,Nair12,Giostri12,SCS12}). 

\begin{figure}
\includegraphics[width=0.5\textwidth]{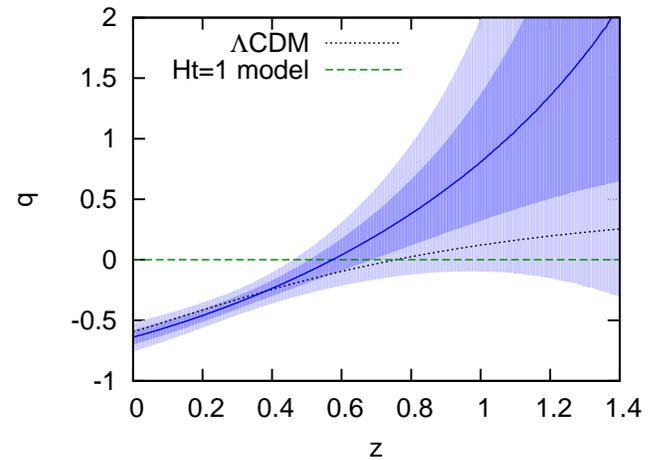}
\caption{\label{Fig:Union_q}
Deceleration parameter estimated from the Union2.1 SN data set. The dark
blue line is the non-parametric smooth reconstruction based on the Gaussian process,
with 68\% and 95\% confidence levels in lighter
blue. The dashed green line presents the prediction of the \Ht, whereas
the dotted black line is for \LCDM.} 
\end{figure}

As an example, we show the deceleration parameter $q(z)$ obtained by
applying Gaussian processes -- a non-parametric and fully Bayesian
approach for data smoothing -- to the distance data given in the
Union2.1 set \citep{Union2.1}.  This result is taken from
\citet{SCS12}, where the recently developed GaPP\footnote{GaPP is
  available for download
  at\\ \texttt{http://www.acgc.uct.ac.za/$\sim$seikel/GAPP/index.html}}
(Gaussian Processes in Python) package is used for the analysis.
\citet{SCS12} also contains an introduction to Gaussian processes.
Here, we have added the line $q\equiv 0$ representing the \Ht{} for
comparison (see Figure \ref{Fig:Union_q}). While the $\Lambda$CDM curve is
well within the 95\% confidence level of the reconstructed $q(z)$, the
\Ht{} is clearly inconsistent with the data.

\section{Hubble rate data}
We now proceed to an analysis based on observational Hubble parameter data
compiled from several sources, independent of SNeIa. We combine measurements obtained with
two methods: cosmic chronometers (CC), which are mainly early type
and/or passively evolving galaxies, and radial baryon acoustic
oscillations (BAO) from galaxy clustering in redshift surveys. Using
these probes should guarantee robustness of our results: possible
systematics are expected to be different in these two approaches and distinct from those in SNeIa analyses.

Based on our dataset, shortly described later in this Section, we will
estimate the Hubble rate and its derivatives with respect to redshift.
For the purpose of our analysis, we will normalise the Hubble
parameter by its current value: $h(z)\equiv H(z)/H_0$. We use
$H_0=70.4\pm2.5\kmsMpc$ \citep{Komatsu11} throughout, although
changing it to $H_0=74.2\pm3.6\kmsMpc$ \citep{Riess09}, preferred by
MS12, would not influence our conclusions.  Subsequently, we use these
estimates to determine $q(z)$ and and the so-called $\Om$
diagnostic. These tests were recently applied by \citet{SYMC12} against the standard model and the results were
consistent with \LCDM. Here we will use the same methods and (slightly
extended) data set to test the \Ht. Before we present the data used,
let us start with the theoretical framework.

In the \Ht, because $a(t)=H_0\, t$ and owing to the general relation
$a(z)=(1+z)^{-1}$ (scale factor is normalised to unity at $t_0$), it
follows that
\beq\label{eq:h's in Ht1}
[H\,t=1]\quad h(z)=1+z\,;\quad h'(z)=1\,;\quad h''(z)=0\;.
\eeq
Finally,
the $\Om^{(1)}$ diagnostic, first introduced by \citet{SSS08}, is defined as
\beq
\Omz\equiv \frac{h^2(z)-1}{(1 + z)^3 - 1}\;.
\eeq
This quantity was proposed to test flat \LCDM{} models because 
in that case it is always equal to the today's value of
the matter density parameter $\Omm(t_0)$. In the \Ht, $\Om^{(1)}$ is
variable and depends on redshift only:
\beq\label{eq:Om in Ht1}
[H\,t=1]\qquad\Omz= \frac{2+z}{3 + 3 z + z^2}\;.
\eeq
In particular, the \Ht{} predicts that $\Om^{(1)}(t_0)=2/3$.

\begin{figure}
\includegraphics[width=0.5\textwidth]{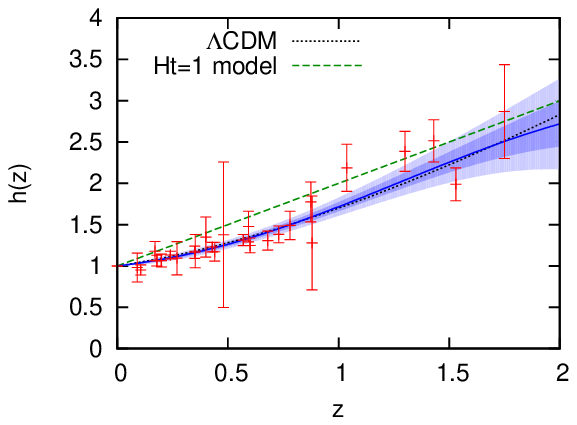}
\includegraphics[width=0.5\textwidth]{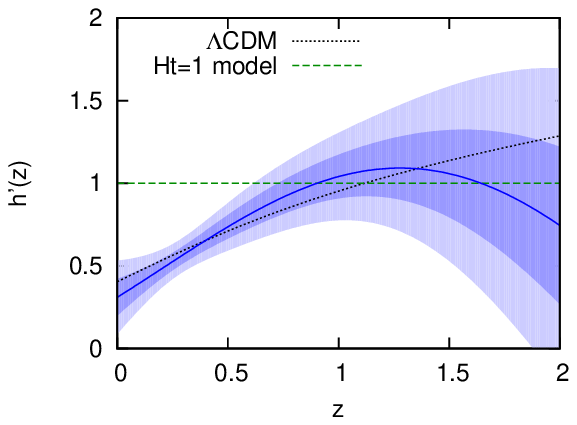} 
\includegraphics[width=0.5\textwidth]{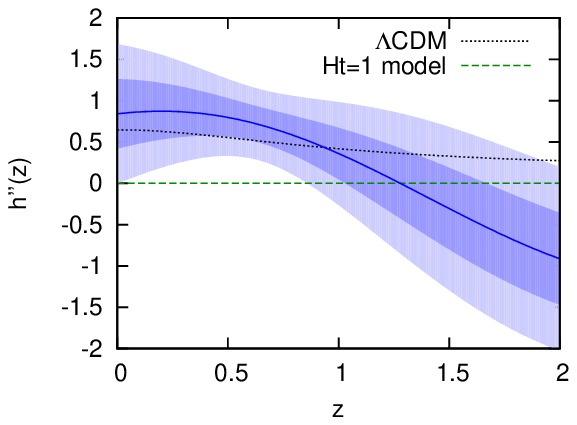} 
\caption{\label{Fig: h h' h'' CC+BAO}
Top to bottom: normalised Hubble parameter, $h(z)\equiv H(z)/H_0$, and
its two derivatives with respect to redshift. Top panel shows data
points from cosmic chronometers and BAO (red crosses with
errorbars). In all the panels, the dark blue lines are the non-parametric
smooth reconstructions of the data, with 68\% and 95\% confidence
levels in lighter blue; the dashed green lines present the predictions of
the \Ht, whereas the dotted black lines are for \LCDM.}
\end{figure}

In order to estimate the above-described quantities  from
observations, we will use the same data as \citet{SYMC12}, with two data
points added. The data set consists of 26 measurements of the Hubble
parameter from the following sources:
\begin{itemize}
\item cosmic chronometers (CC): 18 data points compiled by
  \citet{MVPJC12} from \citet{SVJ05}, \citet{Stern10} and
  \citet{Moresco12}, spanning a redshift interval of $0.09\leq z \leq
  1.75$;
\item radial BAO measurements: 8 data points compiled from
  \citet{GCH09}, \citet{Beutler11}, \citet{ChWa11},  \citet{Blake12}
  and \citet{Reid12}, covering redshifts of $0.106 \leq z \leq 0.73$.
\end{itemize}

\begin{figure}
\includegraphics[width=0.5\textwidth]{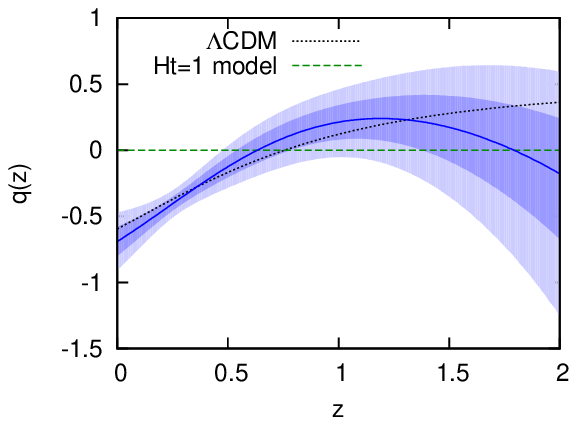}
\includegraphics[width=0.5\textwidth]{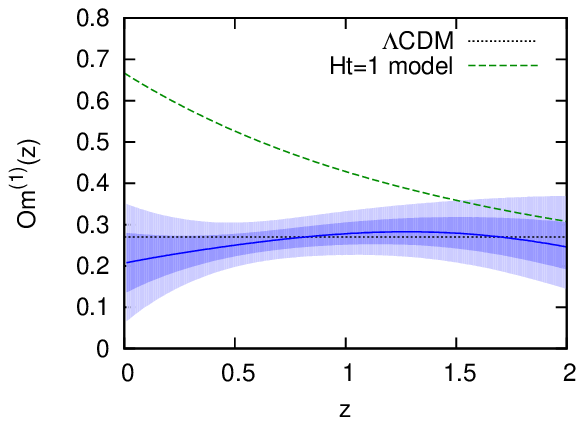} 
\caption{\label{Fig: Om(z) CC+BAO}
The deceleration parameter $q(z)$ (top) and the $\Om^{(1)}$ diagnostic
of cosmic acceleration (bottom). The dark blue line is
the non-parametric smooth reconstruction based on the CC+BAO data,
with 68\% and 95\% confidence levels in lighter blue. The dashed green
line presents the prediction of the \Ht, whereas the dotted black line
is for \LCDM.}
\end{figure}

For the reconstruction of the Hubble rate and its derivatives, we use
again the Gaussian process package GaPP. Such an analysis has already
been done in \citet{SYMC12}. As we use practically the same sample (with only
two BAO measurements added), our findings concerning the
\textit{observed} $h(z)$ etc.\ cannot be much different. However,
unlike in \citet{SYMC12}, here we make comparisons with the \Ht, which
predicts particularly simple forms of the discussed functions, as
given above in Equations (\ref{eq:q in Ht1}), (\ref{eq:h's in Ht1}) and
(\ref{eq:Om in Ht1}) [recall that $q(z)\equiv0$ in the \Ht]. The comparisons are illustrated in Figures
\ref{Fig: h h' h'' CC+BAO} and \ref{Fig: Om(z) CC+BAO}, presenting
respectively $h(z)$, $h'(z)$, $h''(z)$, $q(z)$ and $\Omz$. For
reference, we also plot the curves for the standard flat \LCDM{} model
with $\Omm=0.275$ \citep{Komatsu11}.

The comparison with observational reconstructions is striking. The
predictions of the \Ht{} are in clear disagreement with observations,
especially at low redshifts. Very strong constraints are obtained here
from the first derivative of the Hubble parameter. The \Ht{}
requirement that $h'(z)\equiv1$ is several sigma off the data at
redshifts $z<0.5$.  The same conclusion can be made from the
reconstruction of $q(z)$. A coasting Universe, which is predicted by
the \Ht{}, is inconsistent with observations. The discrepancy is even more pronounced in the
case of the $\Omz$ diagnostic (Figure \ref{Fig: Om(z) CC+BAO}). Here,
the \Ht{} predicts much larger values than those observed, up to
$z\la1.5$. In our opinion, this single plot could be sufficient to
decisively falsify the \Ht.

\section{Summary and conclusions}
\begin{itemize}

\item We analysed some of the predictions of the cosmological model
  proposed by \citet{MeSh12}, in which the scale factor growths
  linearly with time and the Hubble parameter is always the inverse of
  the age of the Universe (the `\Ht'). We started by pointing out some
  inconsistencies in the derivation of the model and its bizarre
  consequences for the nature of the Universe.

\item We referred to several analyses of supernova Ia data that are
  all inconsistent with the \Ht. We presented a plot, which compares
  the deceleration parameter predicted by this model [$q(z)\equiv0$] with a
  model-independent reconstruction of $q(z)$ obtained by applying
  Gaussian processes to the Union2.1 set.

\item We used Hubble parameter data obtained from cosmic chronometers
  (18 data points with $0.09\leq z \leq 1.75$), gathered by
  \citet{MVPJC12}, and from baryon acoustic oscillations (8
  measurements in total), combined from several sources, that spanned
  a redshift interval $0.106 \leq z \leq 0.73$. We compared the
  observational data with the predictions of the \Ht{} regarding the
  following parameters: the Hubble parameter $H(z)$, its derivatives
  with respect to redshift $H'(z)$ and $H''(z)$, as well as the
  $\Om(z)$ diagnostic and the deceleration parameter $q(z)$. In order
  to obtain derivatives of functions based on discrete data, we
  applied the recently developed GaPP package \citep{SCS12}.

\item For all these diagnostics we find severe discrepancies between
  the \Ht{} and observations, especially at low redshifts. The most
  pronounced disagreements are for the first derivative of the Hubble
  parameter, the $\Om$ diagnostic and the deceleration parameter. The
  analysed model completely fails to reproduce their observed
  behaviour for $z<0.5$. This strongly suggests that the cosmological
  model of \citet{MeSh12} is \textit{not} a proper
  description of our Universe.

\item The $\Rh = c\,t$ model proposed by \citet{MeSh12} is mathematically simple,
  but it does not describe the Universe we live in.

\end{itemize}

\section*{Acknowledgements}
We would like to thank Fulvio Melia for stimulating discussions and Geraint F.\ Lewis for drawing our attention to two earlier related papers. This work is funded by the NRF (South Africa).

\bibliographystyle{mn2e}
\bibliography{notRhct}

\end{document}